# Note on Needle in a Haystack

J. Robert Burger
California State University Northridge
CSUN Report Number HCEEN006_4A

**Abstract**
Introduced below is a quantum database method, not only for retrieval but also for creation. It uses a particular structure of true's and false's in a state vector of n qubits, permitting up to 2**2**n words, vastly more than for classical bits. Several copies are produced so that later they can be destructively observed and a word determined with high probability. Grover's algorithm is proposed below to read out, nondestructively the unknown contents of a given stored state vector using only one state vector.

## Data Storage in a State Vector of Qubits

Holding data in a quantum system is clearly problematic, since such systems quickly become incoherent, although systems of the future might be designed to be more stable. This paper assumes a quantum system with m identical sets of n qubits, where each set is assumed to form a state vector. Then each state vector is loaded with 1's and 0's with an appropriate normalization factor. For n qubits a state vector has dimension $2^n$ which theoretically holds $2^{2^n}$ binary codes.

An interesting and practical subset of a general state vector is to initialize qubits to combinations of **H |0> = (1 1)'/√2, |1> = (0 1)'**, and **|0> = (1 0)'**. Each state vector will then have a particular structure termed here the *H10* structure**:** For $2^n$ entries, either the pattern repeats in the other half ($2^{n-1}$) entries, or one of the halves is filled with 0s. For example, one may have 0011, 0000/√2 for n = 3, this being **|0> |1> H |0>**. Note that anything similar to 0111, 0000 is impossible under H10 symmetry, since this rule also applies to each quarter (block of $2^{n-2}$) and so on down to blocks of $2^1$ states. If any one state is true and all others false, the result is just a binary count of n items, a given combination of n computational basis qubits.

Readout is probabilistic. For example, if copies of 0011, 0000 are read a few times, there is a very good chance that entries **|0>, |1>, |0>** and **|1), |1>, |1>** will result, permitting reconstruction of the stored word 0011, 0000. (Simply note that the entries corresponding to combinations of computational basis qubits count up in binary from the left). The H10 structure has the advantage that it can be produced without complex quantum gates. It should be mentioned that the number of codes using H10 symmetry is far more than $2^n$ but still far less than the $2^{2^n}$ available by arbitrarily placing true's and false's into each state of a state vector[1]. Stored words in the proposed system cannot be searched before reading. Ideally Grover's algorithm may be used someday to address, or quickly locate state vectors prior to reading, and then to read them without destroying them.

**Connection to Grover's Algorithm**
One purpose of Grover's quantum algorithm is to find a code that satisfies a given binary function; it solves a satisfiability problem [1, 2]. It should be noted that Grover's quantum algorithm absolutely requires a function to operate. Presumably a function is derived or coded in a form in which it is difficult to locate what satisfies it, since otherwise quantum search would be unnecessary. It would be nonsensical to maintain a classical database of functions just for Grover's algorithm, so a function needs to be constructed on the fly in response to a query.

In the case of an unknown state vector filled with (normalized) 1's and 0's, most likely the function has been forgotten, but it might be possible to reproduce it with quantum gates. Where there is a 1 in the stored state vector, a minus sign must somehow be applied to the corresponding state in a randomized state vector. Assuming this is done (not done here), Grover's algorithm will reproduce with good probability the stored state vector, without damaging it, giving a result similar to the destructive readout of many copies of the state vector directly.

Grover's algorithm applied in a quantum computer will identify the associated code with good probability as long as $M$, the number of ones is less than $N/2$ or $2^{n-1}$. For $1 \leq M \leq N$ the algorithm may employ two extra (dummy) qubits. So for this application $\frac{\pi}{2}\sqrt{\frac{N}{M}}$ iterations of the algorithm are expected. The processed qubits are observed several times until all state vector entries are known with high probability.

**Conclusion**
A method is given above for quantum data creation and subsequent retrieval. It generally avoids the difficult problem of function generation as required for Grover's search, in which, on the fly, a given query must translate to a function, one that is satisfied by a particular set of computational basis qubits (the pattern of which must represent useful data).

Proposed above is a state vector that holds words of data. The data may have H10 symmetry, defined above, which is convenient; but in general a state vector may be filled with any random assortment of 1's and 0's (assuming appropriate normalization).

---

[1] The number of codes with H10 symmetry is $\sum_{i=0}^{n} C(n,i)2^{n-i}$ where C(n, i) are the combinations of i instances of **H|0>** in n qubits.

Several copies may be created. Qubits representing a given word may be observed directly to determine the word with good probability, although once observed, the original qubits are modified. Alternatively, a stored word may be converted (internally) into a quantum function. Then Grover's algorithm may be applied to determine with good probability the stored word without destroying the contents of the state vector.